\documentclass[draft,grl]{agutex}
\usepackage{lineno}
\usepackage{lscape}
\usepackage{graphicx}
\setkeys{Gin}{draft=false}
\authorrunninghead{VALENCIA-CARDONA ET AL.}
\titlerunninghead{Influence of the iron spin crossover in  ... }

\authoraddr{Corresponding author: Juan J. Valencia-Cardona,
Scientific Computing Program, University of Minnesota, Minneapolis, Minnesota, USA.
(valen167@umn.edu)}

\begin{document}

\title{Influence of the iron spin crossover in ferropericlase on the lower mantle geotherm}

\authors{Juan J. Valencia-Cardona,\altaffilmark{1} Gaurav Shukla,\altaffilmark{2} Zhongqing Wu \altaffilmark{3}, Christine Houser \altaffilmark{4},  David A. Yuen \altaffilmark{1,5},  Renata M. Wentzcovitch,\altaffilmark{1,2}}

\altaffiltext{1}{Scientific Computing Program, University of Minnesota, Minneapolis, Minnesota, USA}
\altaffiltext{2}{Department of Chemical Engineering and Materials Science, University of Minnesota, Minneapolis, Minnesota, USA}
\altaffiltext{3}{School of Earth and Space Sciences, University of Science and Technology of China, Hefei, Anhui, China}
\altaffiltext{4}{Earth-Life Science Institute, Tokyo Institute of Technology, Tokyo, Japan}
\altaffiltext{5}{Department of Earth Sciences, University of Minnesota, Minneapolis, USA}

\date{\today}

\clearpage

\begin{abstract}

The iron spin crossover in ferropericlase introduces anomalies in its thermodynamics and
thermoelastic properties. Here we investigate how these anomalies can affect the lower mantle
geotherm. The effect is examined in mantle aggregates consisting of mixtures of bridgmanite,
ferropericlase, and CaSiO$_3$ perovskite, with different Mg/Si ratios varying from harzburgitic to
perovskitic (Mg/Si$\sim$1.5 to 0.8).  We find that the anomalies introduced by the spin crossover increase the
isentropic gradient and thus the geotherm proportionally to the amount of ferropericlase. The
geotherms can be as much as $\sim$ 200K hotter than the conventional adiabatic geotherm at deep lower
mantle conditions. Aggregate elastic moduli and seismic velocities are also sensitive to the spin
crossover and the geotherm, which impacts analyses of lower mantle velocities and composition.

\end{abstract}

\begin{article}

\section{Introduction} \label{sec:intro}

One of the grand challenges in geophysics is to resolve the thermal structure of the Earth's interior. This is clearly not an isolated problem but a fundamental one to clarify the dynamics, evolution, and chemical stratification of the planet \citep{Murakami2}. Besides, a one dimensional (1D) temperature profile is an abstract construct -- a spherically averaged reference temperature model consistent with spherically averaged velocity and composition profiles. To date, numerous one dimensional temperature profiles, or geotherms, have been suggested and calculated by various means and using different assumptions. For example, these include: seismological observations \citep{Brown}, geochemical input \citep{Anderson, Anzellini, Boehler}, mineral physics computations\citep{dasilva, Karki, Wang} or measurements \citep{ Stixrude2}, geodynamic simulations \citep{Matyska}, or a combination of them \citep{Hernlund}. Differences between them arise not only from the technique or input data, but from the constraints to which they are subjected to, i.e., the melting temperature of iron, phase transitions, seismic discontinuities, convection processes, and lower mantle composition. In addition, lateral velocity heterogeneities point to lateral temperature and/or composition variations, a very difficult problem that still awaits, e.g., advances in geodynamic simulations. Thus, construction of one dimensional temperature profiles must be seen as an insufficient but a necessary exercise to advance this topic.

An essential aspect in constructing a geotherm, is to define a suitable potential temperature, i.e., the boundary condition for integration of the adiabatic gradient. Most geotherms are anchored to depths associated with seismic discontinuities (e.g. the $\sim$ 660 km discontinuity), where phase transitions occur \citep{poirier}. For the top of the lower mantle, we assumed a temperature of 1873K at 23GPa as in \cite{Brown} (B\&S) (See also \citep{Akaogi}). The latter is an adiabatic temperature profile constructed from the Debye entropy formulation \citep{Brillouin} and the acoustic velocities from the preliminary reference Earth model (PREM) \citep{Dziewonski}. This geotherm is considered by many to be the standard adiabatic geotherm for the lower mantle. It's important to point out that the conditions at which ringwoodite dissociates into bdg and fp are still somewhat controversial \citep{Irifune3, Chudinovskikh, Katsura}. Thus, the appropriate potential temperature for the lower mantle geotherm is debatable. Several other geotherms, e.g., \cite{Anderson}, \cite{Boehler} and \cite{Anzellini} were obtained by extrapolating the temperature from the inner and outer core geotherm using the phase diagram of iron. Additionally, geotherms constructed from first- principles include diffferent approaches; for instance, \cite{dasilva} and \cite{Karki} calculated the temperatures needed to fit the bulk modulus of pyrolite to PREM. Others such as \cite{Tsuchiya16}, constructed the geotherm from the set of temperatures of different isobars at which the vibrational entropy of bridgmanite was constant. 

In this work, we integrated the isentropic gradient formula using thermodynamics properties of minerals obtained by first principles calculations, starting from the standard boundary condition at 660 km depth, i.e., the experimentally determined post-spinel transition conditions \citep{Akaogi}, also used by Brown and Shankland \citep{Brown}. The relevant lower mantle phases are bridgmanite (Al- Fe- bearing MgSiO$_3$ pervoskite) (bdg), CaSiO$_3$ perovskite (CaPv), and (Mg,Fe)O ferropericlase (fp) \citep{ Irifune2,Irifune, Murakami2}. These minerals form a variety of aggregates commonly characterized by their Mg/Si molar ratio. However, the relative abundances of these aggregates in the lower mantle is still debatable \citep{Irifune2,Irifune,Murakami2, Wu2,Wang, Wu3}. Here we derived isentropes for likely mantle aggregates such as harzburgite (Mg/Si $\sim$ 1.56) \citep{baker}, chondrite (Mg/Si $\sim$ 1.07) \citep{Hart}, pyrolite (Mg/Si $\sim$ 1.24) \citep{McDonough}, peridotite (Mg/Si $\sim$ 1.30) \citep{hirose}, and perovskite only (Mg/Si $\sim$ 0.82) \citep{Williams} to asses the effect of Mg/Si ratio on the isentrope. The presence of FeO in these aggregates needs special consideration. It is well known that ferrous iron (Fe$^{+2}$) in fp exhibits a spin crossover at lower mantle conditions \citep{Badro2003,Speziale,Tsuchiya06}, which introduces anomalies in its thermodynamics \citep{Wentzcovitch09, Wu4} and thermoelastic properties \citep{Crowhurst,Marquardt,Antonangeli,Murakami2,Wu1,Wu2}. Here we investigated in details how these anomalies affect the isentropes of these aggregates in the lower mantle. We compare the temperature gradients of aggregates with iron in fp in the high spin (HS) state and iron undergoing the spin crossover, i.e., in a mixed spin state (MS) of HS and low spin (LS). Clarification of the effect of this spin crossover in the isentropes of several aggregates is a first order question in advancing the problem of mantle temperatures. Finally, we examined the effect such spin crossover induced thermal anomalies have on aggregate velocities.

\section{Method and Calculation Details} \label{sec:method}

The thermoelastic properties of bdg Mg$_{1-x}$Fe$_{x}$SiO$_3$ ($x = 0$ and $0.125$) and fp Mg$_{1-y}$Fe$_{y}$O ($y = 0$ and $0.1875$) were obtained from \cite{Shukla15a} and \cite{Wu1} respectively. Results for other $x$ and $y$ concentrations shown in this work were linearly interpolated and only high spin (HS) Fe$^{2+}$ -bdg was used since no spin crossover in Fe$^{2+}$ occurs in bdg at lower mantle conditions \citep{Bengston,Hsu0,Hsu, Shukla15a, Shukla15b,Lin2}. For CaPv, thermoelastic properties from \cite{Kawai} were reproduced within the Mie-Debye-Gr\"uneisen  \citep{Stixrude} formalism (see Supporting Information). 

We considered mixtures of SiO$_2$ - MgO - CaO -FeO for relevant mantle aggregates; namely, harzburgite  \citep{baker}, chondrite \citep{Hart}, pyrolite  \citep{McDonough}, peridotite \citep{hirose}, and perovskitic only \citep{Williams}. We imposed a fixed iorn partitioning coefficient $K_D = \frac{x/(1-x)}{y/(1-y)}$ between bdg and fp of 0.5 \citep{Irifune}. This assumption is not questioned in the present work, but should be further investigated theoretically in the future. Also, the effect of Al$_2$O$_3$ was not assessed here. Instead, its moles were equally distributed between SiO$_2$ and MgO to keep the Mg/Si molar ratio almost unchanged. We show in Tables S1 and S2 the weight, molar, and volume percentages of oxides and minerals in the aggregates considered. 

The isentropes of different minerals and aggregates were found from their isentropic gradients computed as, 

\begin{equation}  
\label{eq:agradient}
\left( \frac{\partial T}{\partial P} \right)_S =  \frac{\alpha_{agg} V_{agg} T}{ C_{p_{agg}}}
\end{equation}

where the aggregate quantities  $V_{agg}=\sum_i \phi_iV_i$, $\alpha_{agg}= \sum_i \alpha_i \phi_i V_i / V_{agg} $, and $C_{p_{agg}} = \sum_i \phi_i C_{p_i}$ are the molar volume, thermal expansion coefficient, and isobaric specific heat, respectively. Here, $\phi_i$, $V_i$, $\alpha_i$, and $Cp_i$ represent respectively the molar fraction, molar volume, thermal expansion coefficient, and isobaric specific heat of the i$^{th}$ mineral in the mixture. All the isentropes, here loosely referred as geotherms, were compared with the adiabatic \citep{Brown} and superadiabatic \citep{Anderson} geotherms. Once these geotherms are obtained, we compute aggregate velocities along aggregate specific geotherms using the Voigt-Reuss-Hill (VRH) average of elastic moduli and compared with PREM values \citep{Dziewonski}.

\section{Results and Discussion} \label{sec:result}

\subsection{Lower mantle mineral isentropes}\label {fpgeo}

\begin{figure}[htbp]
\centering
  \resizebox{15cm}{!}{\includegraphics{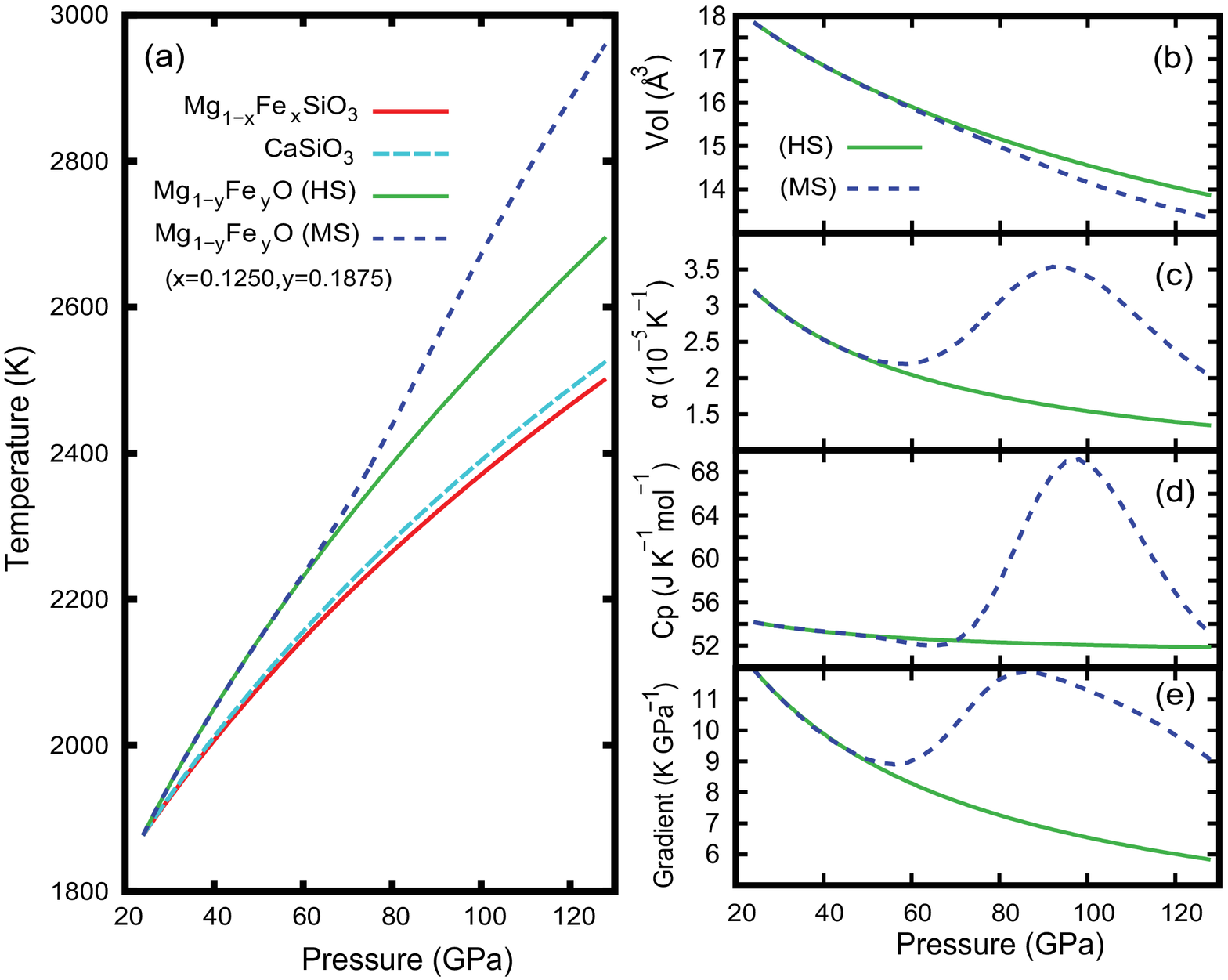}}
	\caption{(a) Isentropes for bdg, CaPv, and fp in HS and MS states. (b) Volume, (c) thermal expansion coefficient ($\alpha$), (d) isobaric specific heat (C$_P$), and (e) adiabatic gradient for fp in HS and MS states.}
	\label{fig:fpgeoth}
\end{figure}

To unravel the possible consequences of the iron spin crossover in fp on the lower mantle geotherm, first we calculated using  Eq.\ref{eq:agradient}, the isentropes of  (Mg$_{0.875}$Fe$_{0.125}$)SiO$_3$, CaSiO$_3$, and (Mg$_{0.8125}$Fe$_{0.1875}$)O with iron in HS and MS states (See Figure \ref{fig:fpgeoth}). Here, differences between (Mg$_{0.875}$Fe$_{0.125}$)SiO$_3$ and CaSiO$_3$ are only noticeable at higher pressures, and no sudden changes in the slopes of such isentropes were observed (See Figure \ref{fig:fpgeoth}a). For fp however, the isentropes of MS and HS states differ by $\sim$ 260 K at deep lower mantle pressures (Figure \ref{fig:fpgeoth}a). The anomalies caused by the spin crossover on $V$,$\alpha$, and $C_p$ (See Figures \ref{fig:fpgeoth}b to \ref{fig:fpgeoth}d) on the adiabatic gradient of fp (Figure \ref{fig:fpgeoth}e) are responsible for such temperature increase. Note that for $\alpha$ and $C_p$, the spin crossover anomalies correspond to broad peaks at similar pressure ranges, but do not cancel during the adiabatic gradient integration owing to significant differences in their magnitudes (Figures \ref{fig:fpgeoth}c and \ref{fig:fpgeoth}d).

\subsection{Effect of Mg/Si ratio fp spin crossover on the geotherm} \label {geo}

\begin{figure}[htbp!]
\centering
  \resizebox{16cm}{!}{\includegraphics{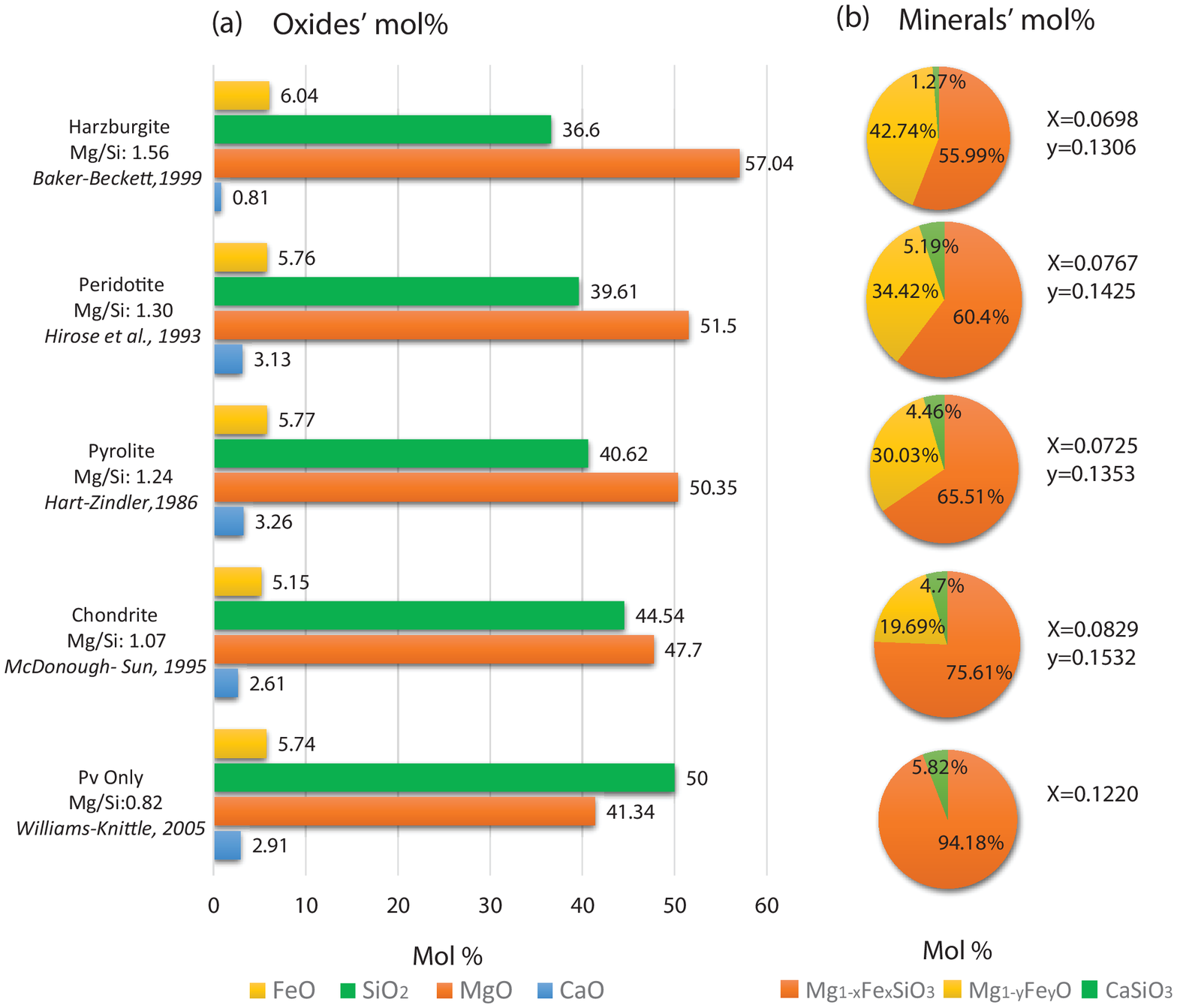}}
  \caption{ Aggregates characterized by their Mg/Si ratio. a) Mol\% of oxides and b) minerals in each aggregate. The x and y values correspond to the molar fraction of iron in bdg and fp respectively. }
	\label{fig:charts}
\end{figure}

We now investigate the isentropes of aggregates likely to be present in the lower mantle. Their compositions are shown in Figure \ref{fig:charts}. The horizontal charts in Figure \ref{fig:charts}a represent the oxides mol\% of SiO$_2$ - MgO - CaO –FeO in the aggregate, while the pie charts in Figure \ref{fig:charts}b show the mol\% of each mineral (bdg,fp,CaPv) in the aggregate. Each composition is characterized by its Mg/Si molar ratio, and it is clear that as Mg/Si decreases, so does the amount of fp. Further details about these compositions can be found in the Supporting Information. 

Al$_2$O$_3$ and Fe$_2$O$_3$ were not regarded here and their inclusion in the lower mantle mixture is important to fully discern the lower mantle thermal behavior and composition. However,  due to the presumptive low ferric iron concentration in the lower mantle (Fe$^{3+}$/$\sum$Fe $\sim$ 0.01-0.07) \citep{Shenzhen}, the spin crossover effect in the (Mg$_{1-y}$Fe$^{3+}_y$)(Si$_{1-y}$Fe$^{3+}_y$)O$_3$ thermodynamic properties and thus the isentrope is negligible \citep{Shukla16b}. Also, it was recently shown by \cite{Shukla2} that the thermodynamic and thermoelastic properties of (Mg$_{1-y}$Fe$^{3+}_y$)(Si$_{1-y}$Al$_y$)O$_3$ are very similar to those of (Mg$_{1-x}$Fe$^{2+}_x$)SiO$_3$, and therefore, the results shown here should not vary significantly. However, the contribution of CaO was not ignored as is usually done in the literature \citep{ Wentzcovitch04,Murakami2,Wang} due to its relatively small abundance in the lower mantle. The thermoelastic properties of CaPv obtained by \cite{Stixrude2} and \cite{Tsuchiya} are quite significant (Figures S5 and S7). The smaller shear modulus of \cite{Tsuchiya} helps to reduce discrepancies with PREM values for all aggregates containing CaPv.

\begin{figure}[htbp]
\centering
  \resizebox{16cm}{!}{\includegraphics{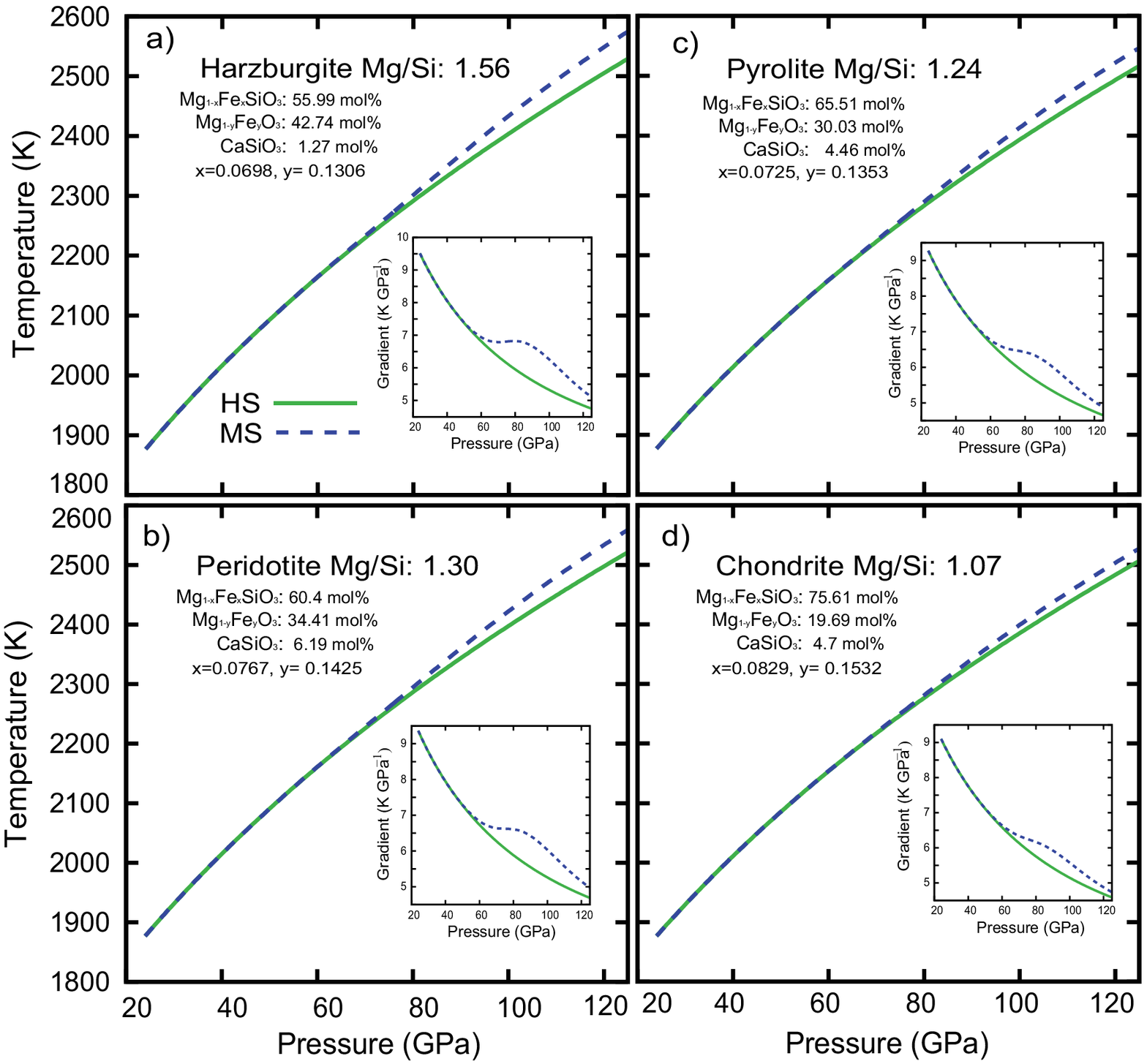}}
	\caption{
	Adiabats for a) harzburgite, b) peridotite, c) pyrolite, and d) chondrite with fp in HS and MS states. The values of x and y correspond to the amount of iron in bdg and fp respectively. Solid/dashed lines correspond to adiabats with fp in HS/MS states. Inset figures show the adiabatic gradient for each aggregate.}
	\label{fig:HS_MS}
\end{figure}

We first examined the temperature increments caused by the spin crossover in all aggregates in which fp does (MS) and does not (HS) undergo spin crossover  (Figure \ref{fig:HS_MS}). Adiabatic geotherms and gradients (insets) for harzburgite, peridotite, pyrolite, and chondrite are shown in Figures \ref{fig:HS_MS}a to \ref{fig:HS_MS}d. The temperature differences between HS and MS at high pressures (P $\sim$ 125GPa) for harzburguite ($\sim$ 50 K) were greater than those for peridotite ($\sim$ 40 K), pyrolite ($\sim$ 30 K), and chondrite ($\sim$ 20 K) owing to harzburgite's higher Mg/Si. We also notice that MS gradients (insets) for all aggregates deviate from HS at pressures of $\sim$ 60 GPa, as a consequence of the spin crossover, and the gradient decreases in proportion to the Mg/Si ratio.

\begin{figure}[htbp!]
\centering
  \resizebox{12cm}{!}{\includegraphics{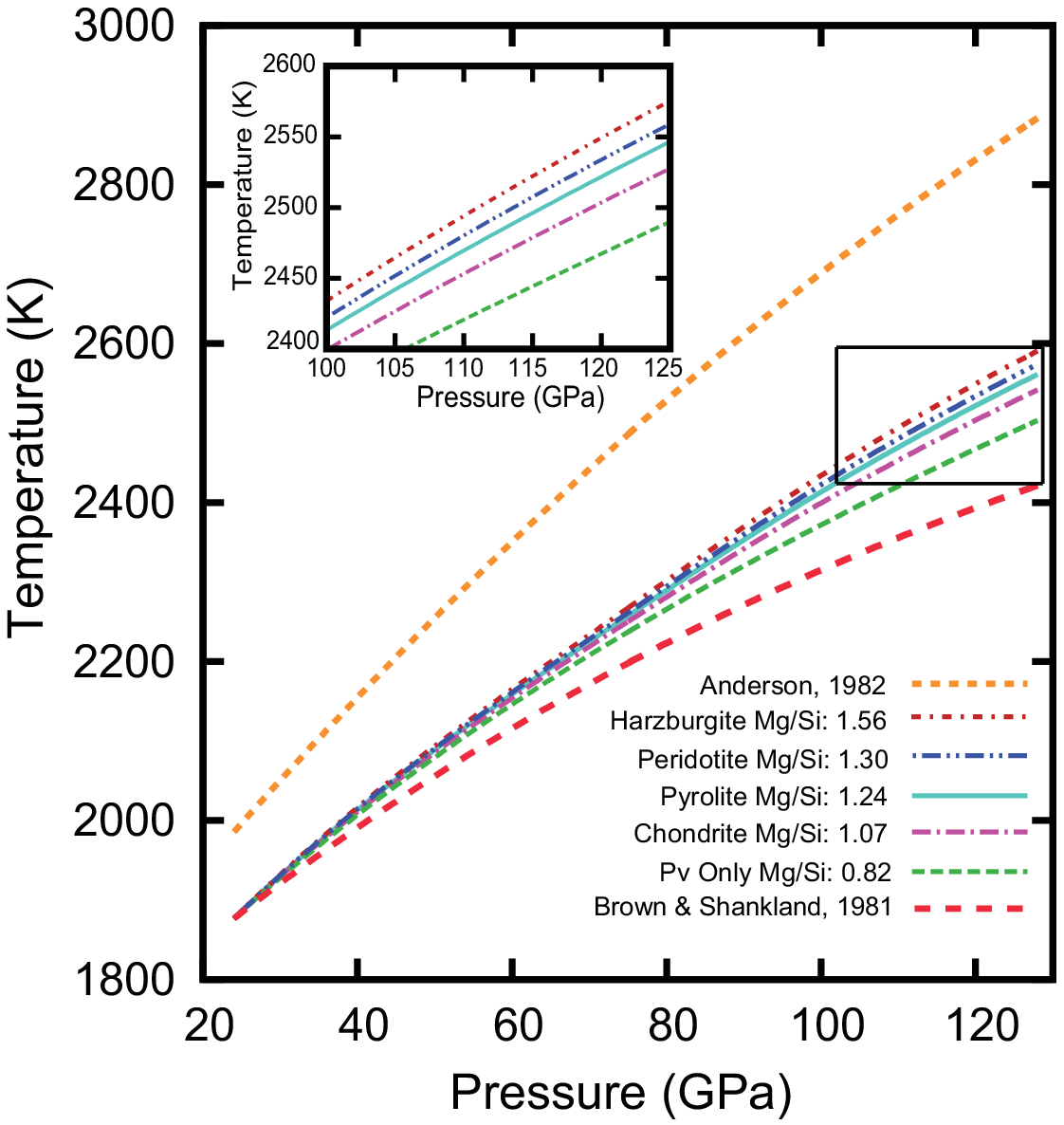}}
  \caption{Calculated adiabats for aggregates considered, compared with geotherms by \cite{Brown} and \cite{Anderson}. Inset corresponds to highlighted region.}
	\label{fig:allgeotherms}
\end{figure}

The isentropes of all aggregates are compared with reference geotherms by \cite{Brown} and \cite{Anderson} in Figure \ref{fig:allgeotherms}. All calculated adiabats were anchored to 1873K at 23GPa as in \cite{Akaogi} and \cite{Brown}. At deep lower mantle conditions, harzburgite achieves the highest temperature, which is about $\sim$ 200 K higher than that of B\&S. Temperatures then decrease with respect to the Mg/Si ratio, i.e., the aggregate's fp fraction. 

\section{Geophysical Significance} \label {Geophys}

\begin{figure}
\centering
\resizebox{14cm}{!}{\includegraphics{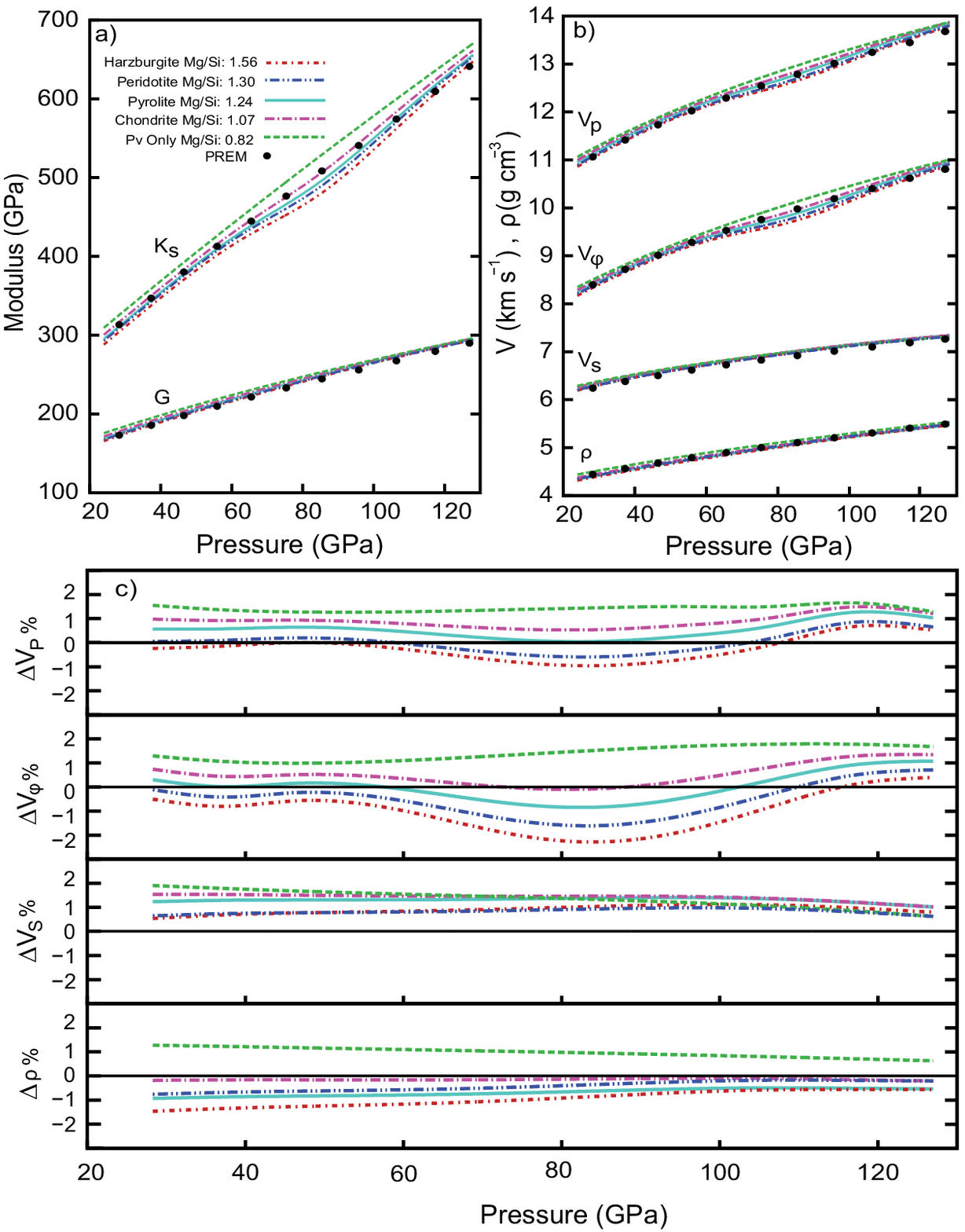}}
\caption{ (a) Elastic moduli ($K_S$, $G$), (b) acoustic velocities ($V_P$, $V_{\phi}$, $V_S$) and densities ($\rho$) for all aggregates considered.  Black circles indicate PREM values \citep{Dziewonski}. (c) Relative deviations from PREM shown as percentages.}
\label{fig:combined}
\end{figure}

The elastic moduli ($K_S$,$G$), acoustic velocities ($V_P$,$V_{\phi}$,$V_S$), and densities ($\rho$) of all the aggregates along their own isentropes are shown in Figure \ref{fig:combined}. In aggregates containing fp, $K_S$ softening due fp spin crossover \citep{Wu1,Wu2} is observed and varies proportionally to the Mg/Si ratio (Figure \ref{fig:combined}a). Moreover, $G$ does not exhibit any anomalous behavior and its value for different compositions converges to similar values at deep lower mantle pressures. For aggregates with higher Mg/Si ratio, calculated $G$ revealed a better agreement  with PREM values, while the $K_S$ values for chondrite and pyrolite were closer to PREM values (Figure \ref{fig:combined}a). The signature of $K_S$ anomalies were also obvious in the compressional ($V_p = \sqrt{( K_S + \frac{4}{3} G )/ \rho} $) and bulk ($V_{\phi} = \sqrt{K_S / \rho}$)  velocities, while shear velocities $V_S = \sqrt{G/\rho}$ were unaffected (See Figure \ref{fig:combined}b). 

The relative velocity deviations ($\Delta V_P$, $\Delta V_{\phi}$, and $\Delta V_S$) from PREM values are shown in Figure \ref{fig:combined}c . $\Delta V_P$ and $\Delta V_{\phi}$ decreased with increasing Mg/Si ratio, and reached negative values mainly due fp spin crossover. However, $\Delta V_S$ for all the aggregates was positive and the deviations from PREM values for perditotite and harzburgite were the lowest. The relative density deviations ($\Delta \rho$) reduced with increasing pressure (Figure \ref{fig:combined}c). For compositions with fp, the volume collapse caused by fp spin crossover \citep{Wentzcovitch09} seemed to reduce such deviations further.

\begin{figure}
\centering
\resizebox{16cm}{!}{\includegraphics{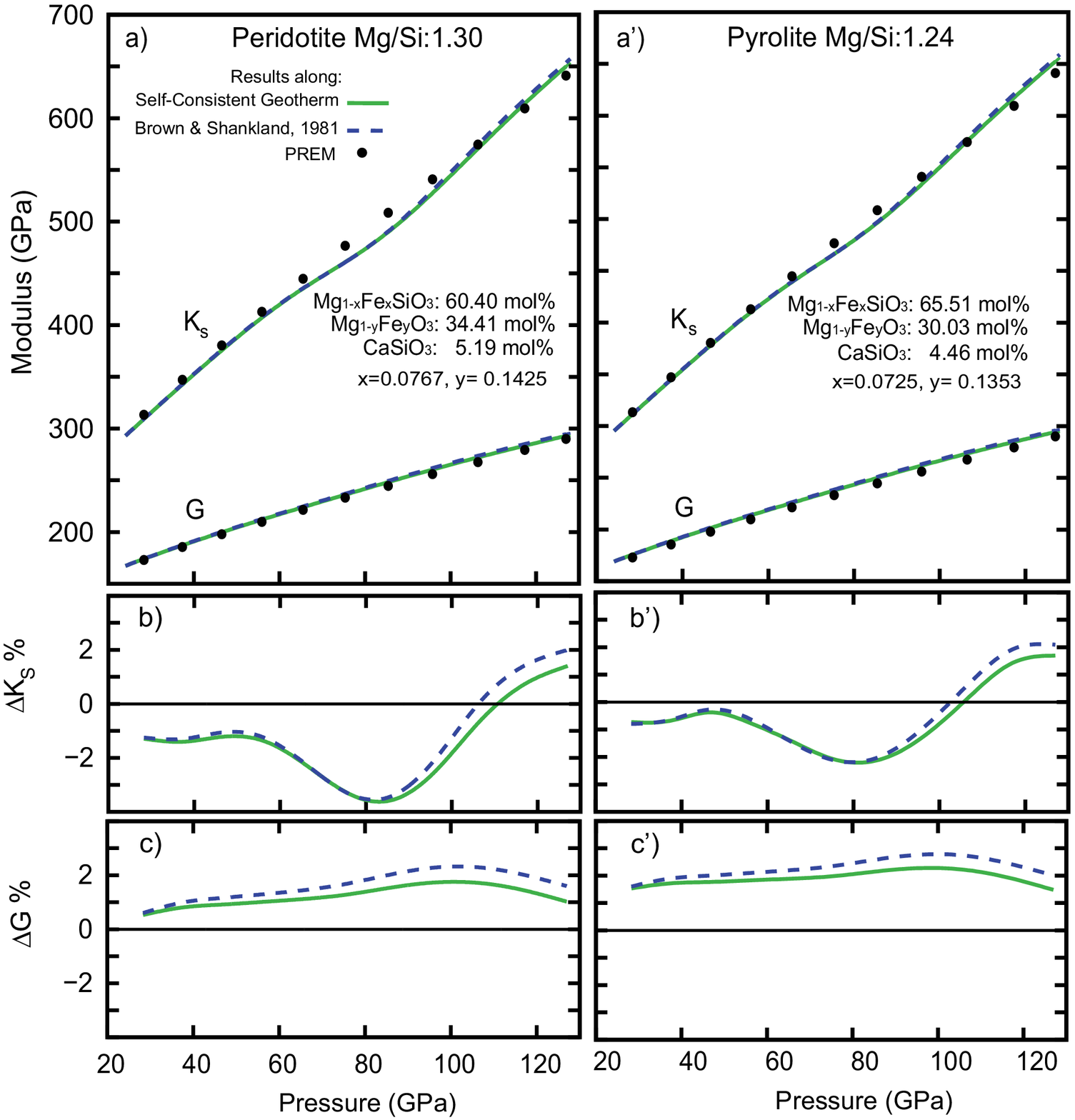}}
\caption{Elastic moduli ($K_S$, $G$) for  (a) peridotite, and (a') pyrolite along self-consistent (solid green line) and \cite{Brown} (dashed blue line) geotherms. Black circles correspond to PREM values \citep{Dziewonski}. (b)/(b')  and (c)/(c') are the elastic moduli relative deviations ($\Delta K_S$, $\Delta G$) with respect to PREM values, shown as percentages for peridotite/pyrolite. }
\label{fig:perido_pyro_interp}
\end{figure}

Moreover, as discussed in section \ref{geo}, the iron spin crossover in fp raises the aggregate's geotherm in proportion to its fp content  (or to the iron content in fp). This temperature boost can affect analyses of the lower mantle composition. Figures \ref{fig:perido_pyro_interp}a and \ref{fig:perido_pyro_interp}a' show peridotite and pyrolite's elastic moduli along their own geotherms and the B\&S geotherm. Although not substantial, relative deviations from PREM varied depending on the geotherm used. For the bulk modulus $\Delta K_S$ in Figures \ref{fig:perido_pyro_interp}b and \ref{fig:perido_pyro_interp}b', the relative deviations from PREM were the same along both geotherms until pressures lower than 80 GPa, while for the shear modulus relative deviations $\Delta G$, results along the self-consistent geotherm (Figures \ref{fig:perido_pyro_interp}c and \ref{fig:perido_pyro_interp}c') exhibited lower deviations from PREM throughout the whole lower mantle. Such details suggest that the aggregate thermoelastic properties along the self-consistent geotherm should sharpen uncertainties in analyses of lower mantle composition.

\section{Conclusions} \label{sec:conc}

We presented a set of adiabatic geotherms for individual minerals and likely lower mantle aggregates under realistic conditions. We showed that the spin crossover in ferropericlase introduces an anomaly in its isentrope similar to ``superadiabaticity''. This effect increases the adiabatic temperature gradient in different aggregates in proportion to their ferropericlase content or Mg/Si ratio. Velocities of aggregates with compositions varying from perovskitic to harzburgitic along their self-consistent geotherms exhibited deviations from PREM velocities within $\pm$ $\sim$ 2\%. However, pyrolitic and peridotitic compositions displayed the best fit with respect to PREM values. The elastic moduli, velocities, and densities of these aggregates along their own self-consistent geotherms tend to display smaller deviations from PREM (up to $\sim$ 1\% less), than those along standard geotherms such as \cite{Brown}. We stress here that we cannot afford to ignore the impact from the spin crossover anomalies, because neglecting them would lead to erroneous interpretations of the lower mantle geotherm and composition.

\begin{acknowledgement}
The authors acknowledge Caroline Qian for her early contributions to the isentrope code. 
This work was supported primarily by grants NSF/EAR 1319368, 1348066, and NSF/CAREER 1151738. Zhongqing Wu was supported by State Key Development Program of Basic Research of China (2014CB845905) and NSF of China (41274087). C. Houser was supported by the Earth-Life Science Institute at Tokyo Institute of Technology. Results produced in this study are available in the supporting information.
\end{acknowledgement}

\section{Supplementary Material}

\noindent\textbf{Introduction}

The supporting information consists of Texts S1-S4, Figures S1-S7, and Tables S1-S4. Figure S1 shows properties of pyrolite  ($V$, $\alpha$, and C$_P$) with fp in HS and MS states. 

Elastic moduli and acoustic velocities for the different aggregates are shown in Figures S2 to S4. Figure S5 shows the isentropes for CaSiO$_3$ using MDG parameters from \cite{Stixrude2} and \cite{Kawai}. Figure S6 shows the elastic constants of CaPv reported by \cite{Tsuchiya} fitted to a Mie-Debye-Gr\"uneisen model. Figure S7 depicts the compression curves and moduli of CaPV using \cite{Stixrude2} and \cite{Kawai,Tsuchiya} MDG parameters. Texts S1-S3 are supporting analyses of what is presented in Figures S1-S7.

Table S1 shows the mol\% and wt\% of the oxides of each of the aggregates considered. It also shows how the compositions were before the Al$_2$O$_3$ moles were adjusted.
Table S2 shows the mol\%, vol\%, and wt\% of each of the minerals in all the aggregates, along with the Fe concentrations in bdg and fp (x and y). Tables S3 and S4 shows the adiabats for the minerals and aggregates studied.

\noindent\textbf{Text S1.}
 We compared the pyrolite's adiabat properties when its fp (30 mol\% in pyrolite) did (MS) and did not (HS) undergo spin transition. Figure S1 shows pyrolite's $\alpha$, $C_p$ and $V$ calculated along its own geotherm. Compression curves differences are subtle (Figure S1a), while the anomalies introduced in  $\alpha$ and $C_p$ (Figures S1b and S1c respectively) cause broad peaks throughout the spin crossover at equivalent pressure ranges.
 
\noindent\textbf{Text S2.}
Elastic moduli, velocities and densities of the minerals in each aggregate, calculated along the aggregate geotherm, can be found in Figures S2, S3, and S4. It can be observed that $K_S$ for bdg and CaPv behave similarly, but CaPv shear modulus $G$ is significantly smaller than that of bdg. 
In all aggregates, $K_S$ for both perovskites is larger than PREM values. $G$ for bdg displays values greater than PREM, while CaPv $G$ agrees well until mid lower mantle pressures, and then deviates to values smaller than those of PREM. Finally, moduli for fp are always smaller than PREM values and showed clear anomalies associated with spin crossover. The velocities followed a trend consistent with that of the moduli, while for the densities we observed that CaPv is denser than bdg. Also, fp increases its density at mid lower mantle pressures due the volume collapse caused by the spin crossover.

\noindent\textbf{Text S3.}
Isentropes for CaPv were calculated using \cite{Stixrude2} MDG parameters, which are different from those obtained by \cite{Kawai} (See Figure S5).  They differ by $\sim$ 200K in the deep lower mantle. Therefore, the thermodynamic model adopted for CaPv has some influence on the geotherm. 

We used CaPv properties reported by \cite{Kawai,Tsuchiya} for all the aggregates. Figure S6 shows the elastic constants of CaPv for different isotherms, while Figures S7a and S7b  compare compression curves and moduli between \cite{Stixrude2} and \cite{Kawai,Tsuchiya}.

\begin{figure}
\setfigurenum{S1} 
\centering
\noindent\includegraphics[width=17cm]{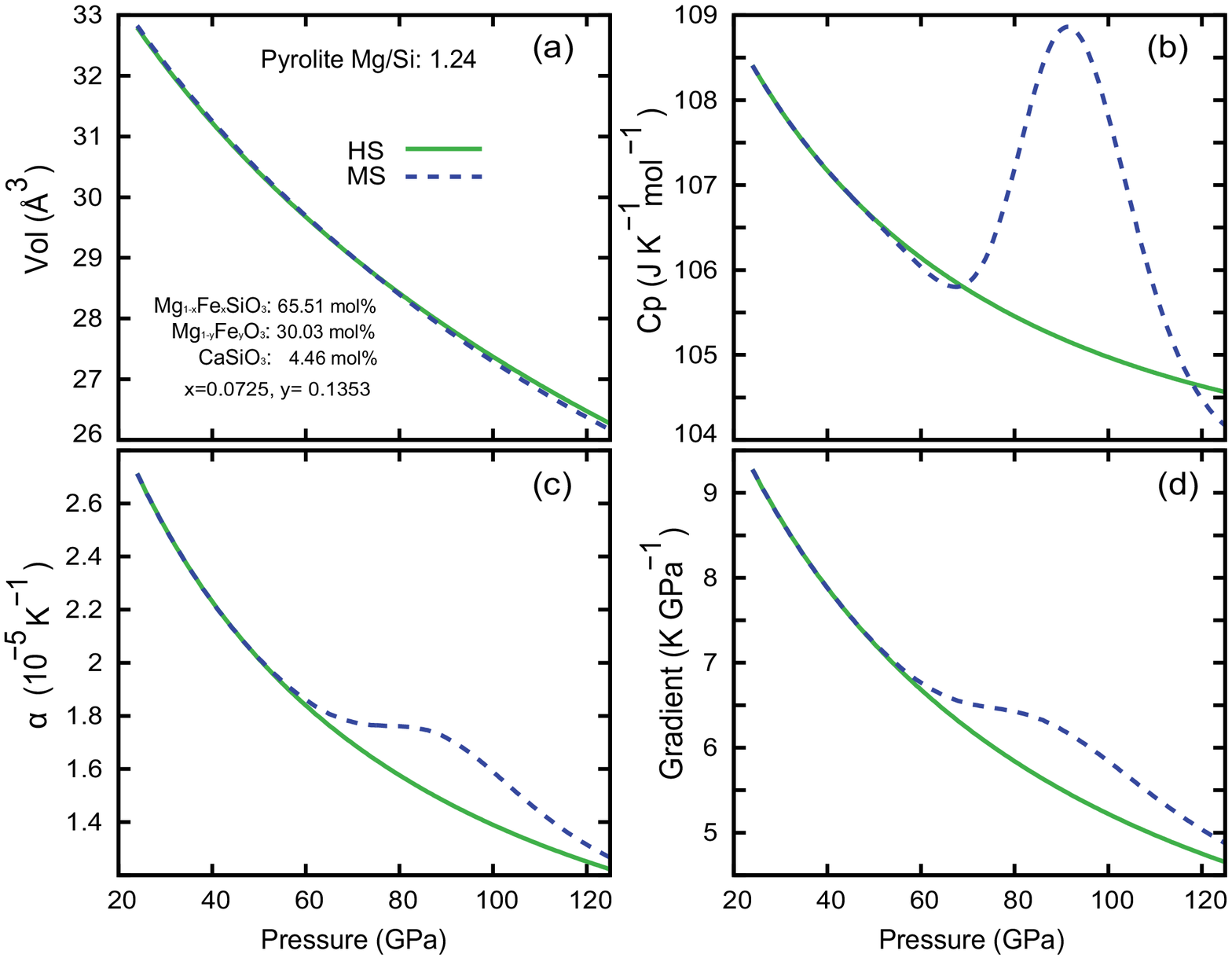}
	\caption{
	Properties of pyrolite with ferropericlase in high spin state (HS) and mixed spin (MS) state along its own adiabat. (a) Volume, (b) thermal expansion coefficient, (c) isobaric specific heat and (d) adiabatic gradient.}
	\label{figure1}
\end{figure}

\begin{figure}
\setfigurenum{S2} 
\centering
\noindent\includegraphics[width=16cm]{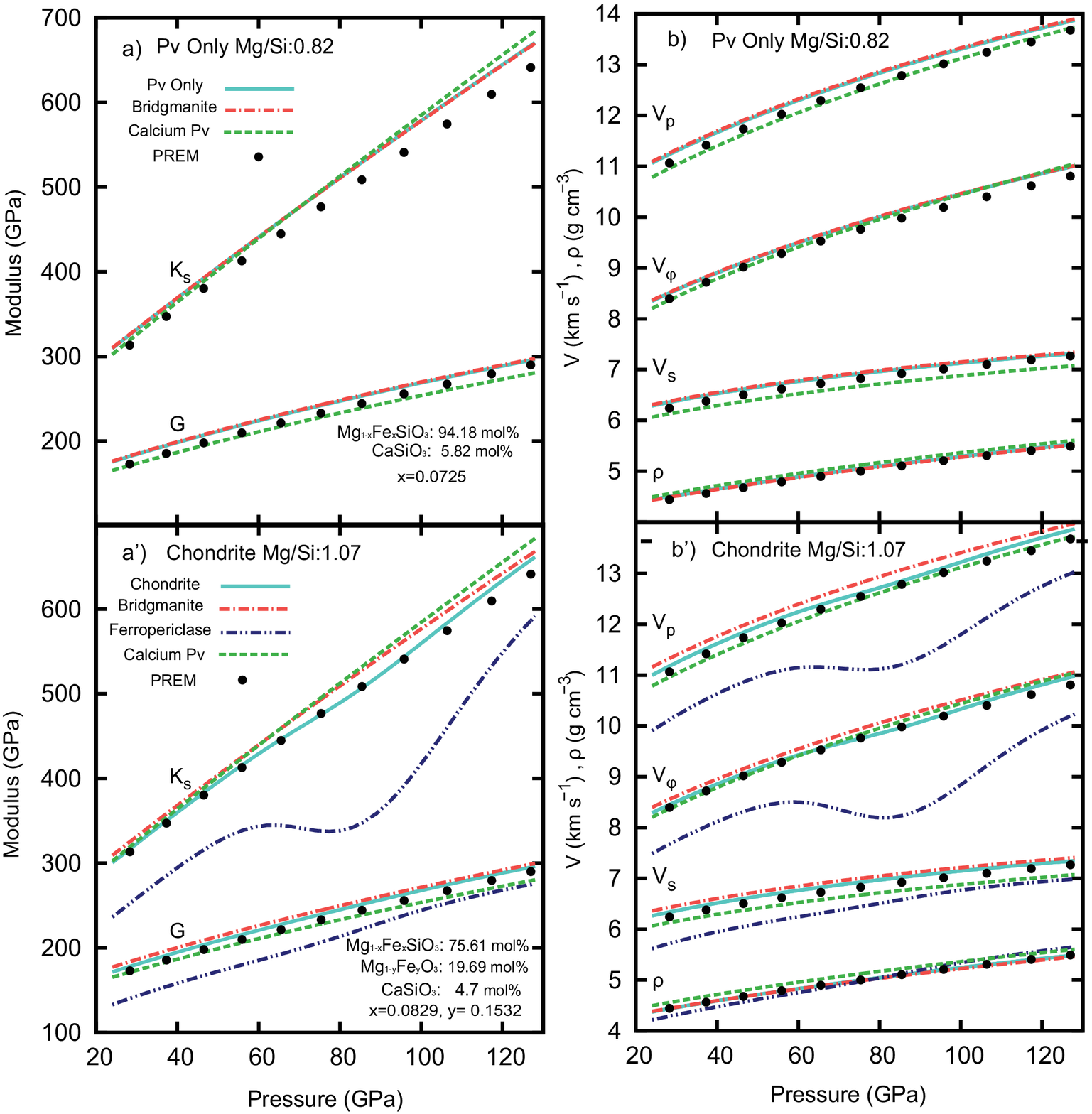}
\caption{(a) and (a') are the elastic moduli ($K_S$, $G$) for Pv Only and chondritic aggregates respectively, while (b) and (b') are the acoustic velocities ($V_P$, $V_{\phi}$, and $V_S$) and densities ($\rho$). The solid/dashed lines represent the aggregate/mineral properties.}
\label{figure2}
\end{figure}

\begin{figure}
\setfigurenum{S3} 
\centering
\noindent\includegraphics[width=16cm]{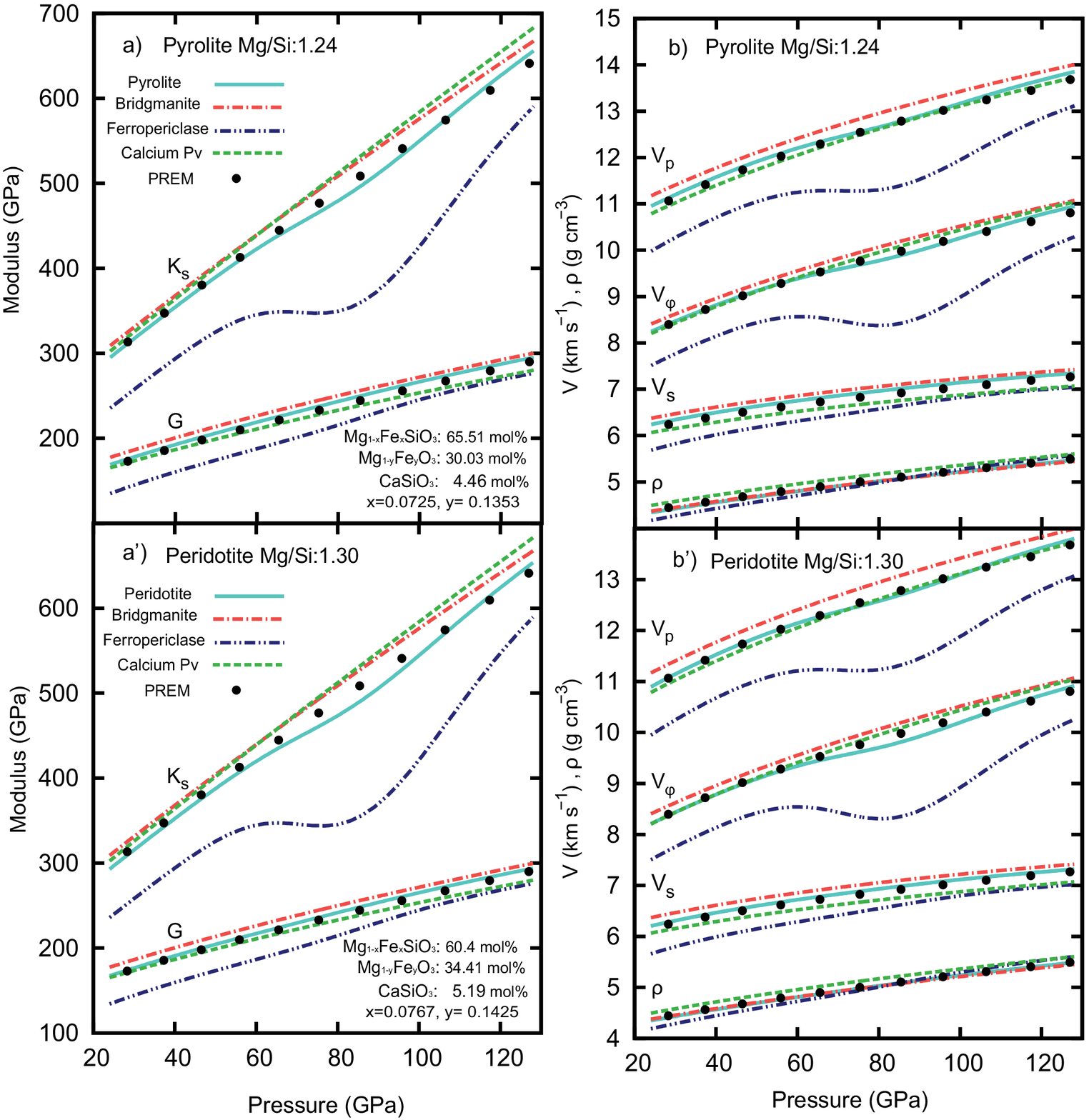}
\caption{(a) and (a') are the elastic moduli ($K_S$, $G$) for pyrolite and peridotite compositions respectively, while (b) and (b') are the acoustic velocities ($V_P$, $V_{\phi}$, and $V_S$)  and densities ($\rho$). The solid/dashed lines represent the aggregate/mineral properties.}
\label{figure3}
\end{figure}

\begin{figure}
\setfigurenum{S4} 
\centering
\noindent\includegraphics[width=16cm]{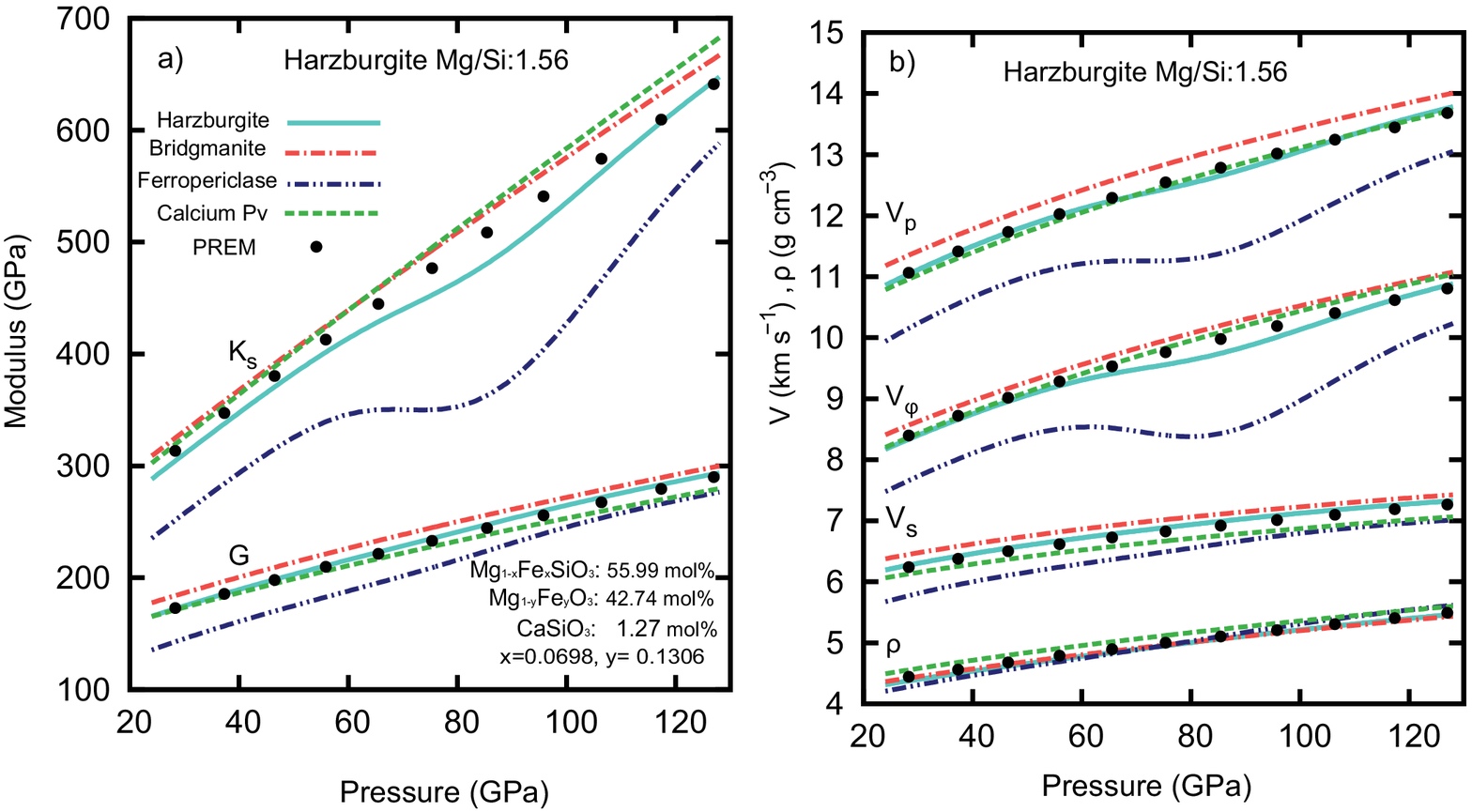}
\caption{(a) Elastic moduli ($K_S$, $G$), (b) acoustic velocities ($V_P$, $V_{\phi}$, and $V_S$), and densities ($\rho$) for harzburgite. The solid/dashed lines represent the aggregate/mineral properties}
\label{figure4}
\end{figure}

\begin{figure}
\setfigurenum{S5} 
\centering
\noindent\includegraphics[width=12cm]{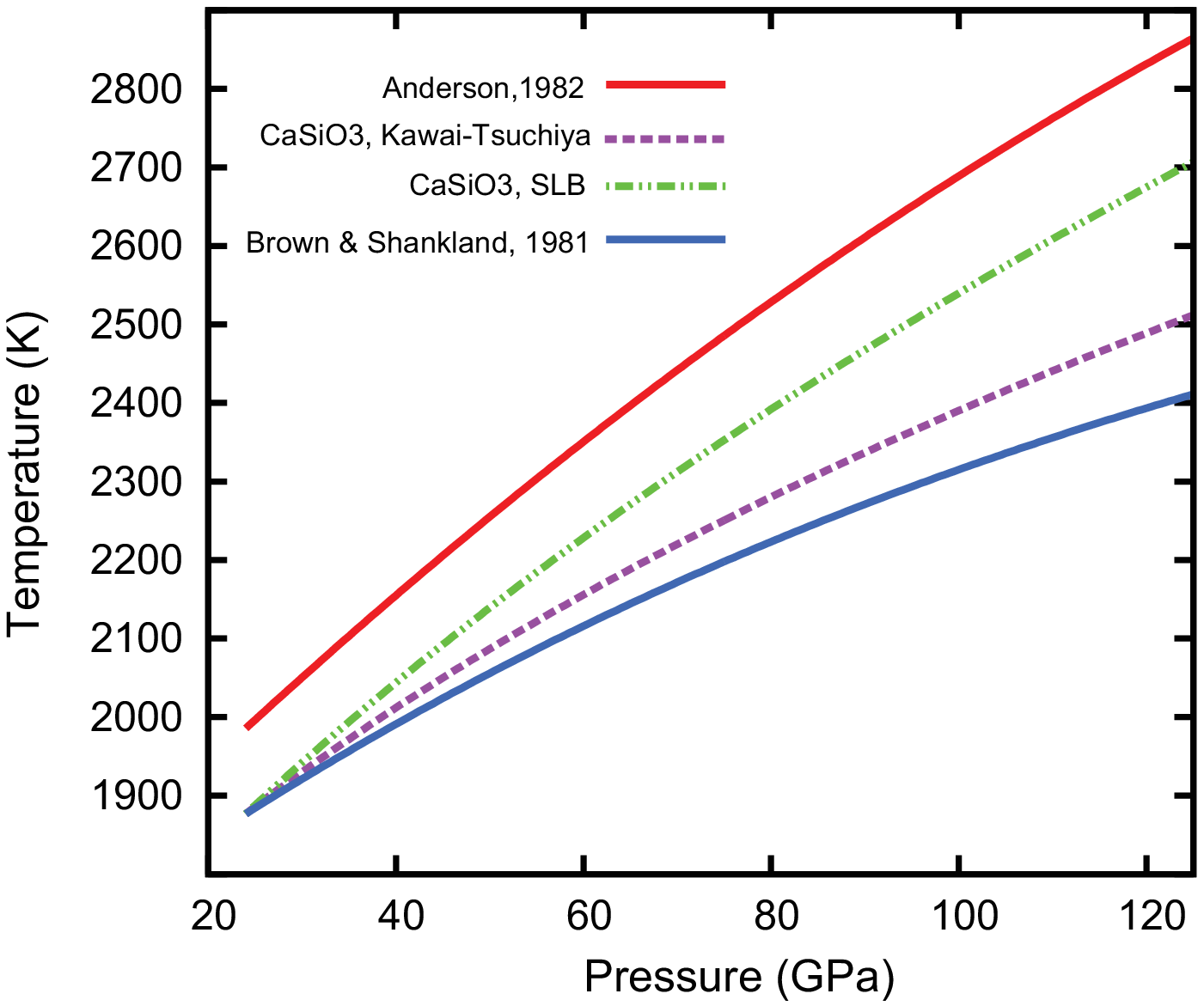}
\caption{Calculated CaSiO$_3$ adiabats using MDG parameters from \cite{Stixrude2} (SLB) and \cite{Kawai}. At deep lower mantle pressures they differ by almost 200K }
\label{figure5}
\end{figure}

\begin{figure}
\setfigurenum{S6} 
\centering
\noindent\includegraphics[width=8cm]{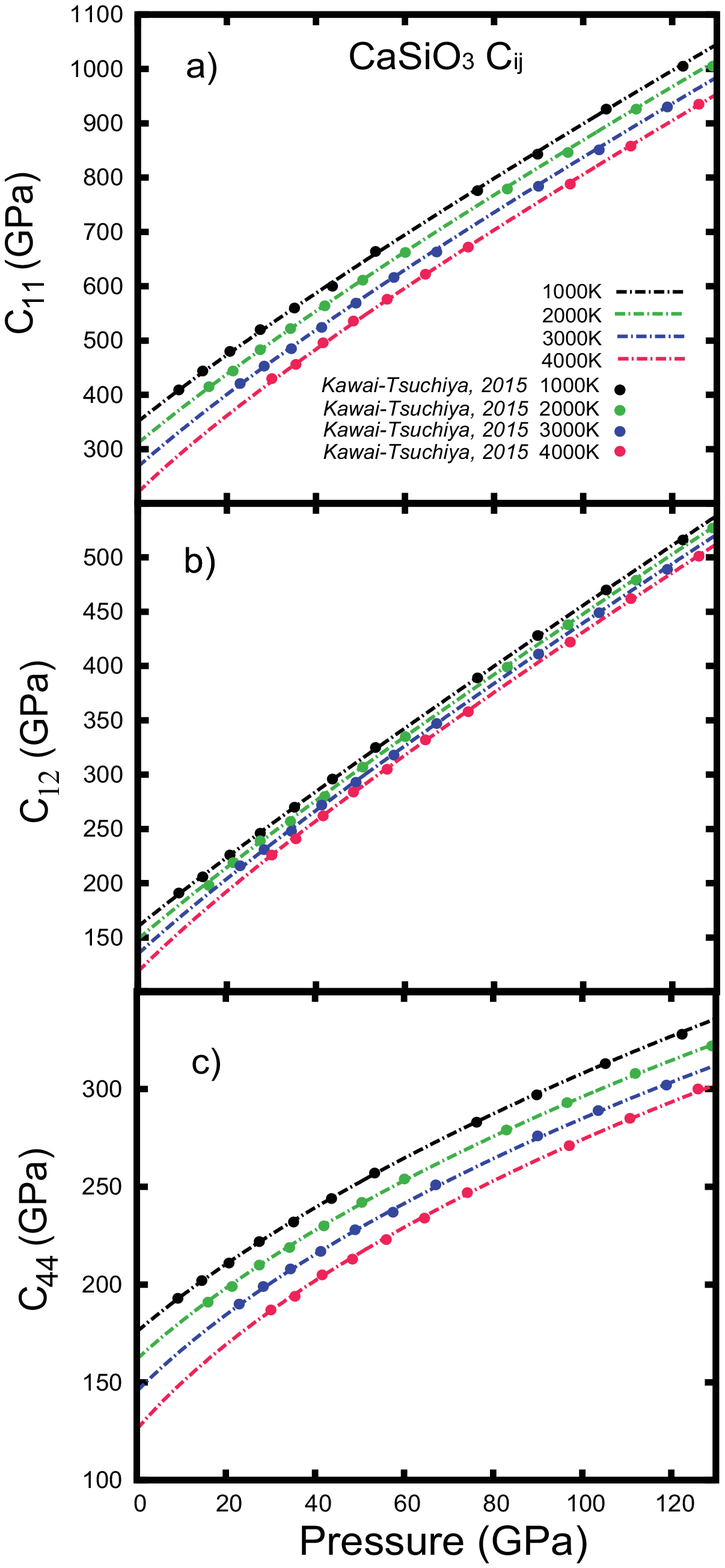}
\caption{Results by \cite{Tsuchiya} fitted to a Mie-Debye-Gr\"uneisen model developed in \cite{Stixrude} ( See Eq. 2 in \cite{Tsuchiya}) }
\label{figure6}
\end{figure}

\begin{figure}
\setfigurenum{S7} 
\centering
\noindent\includegraphics[width=10cm]{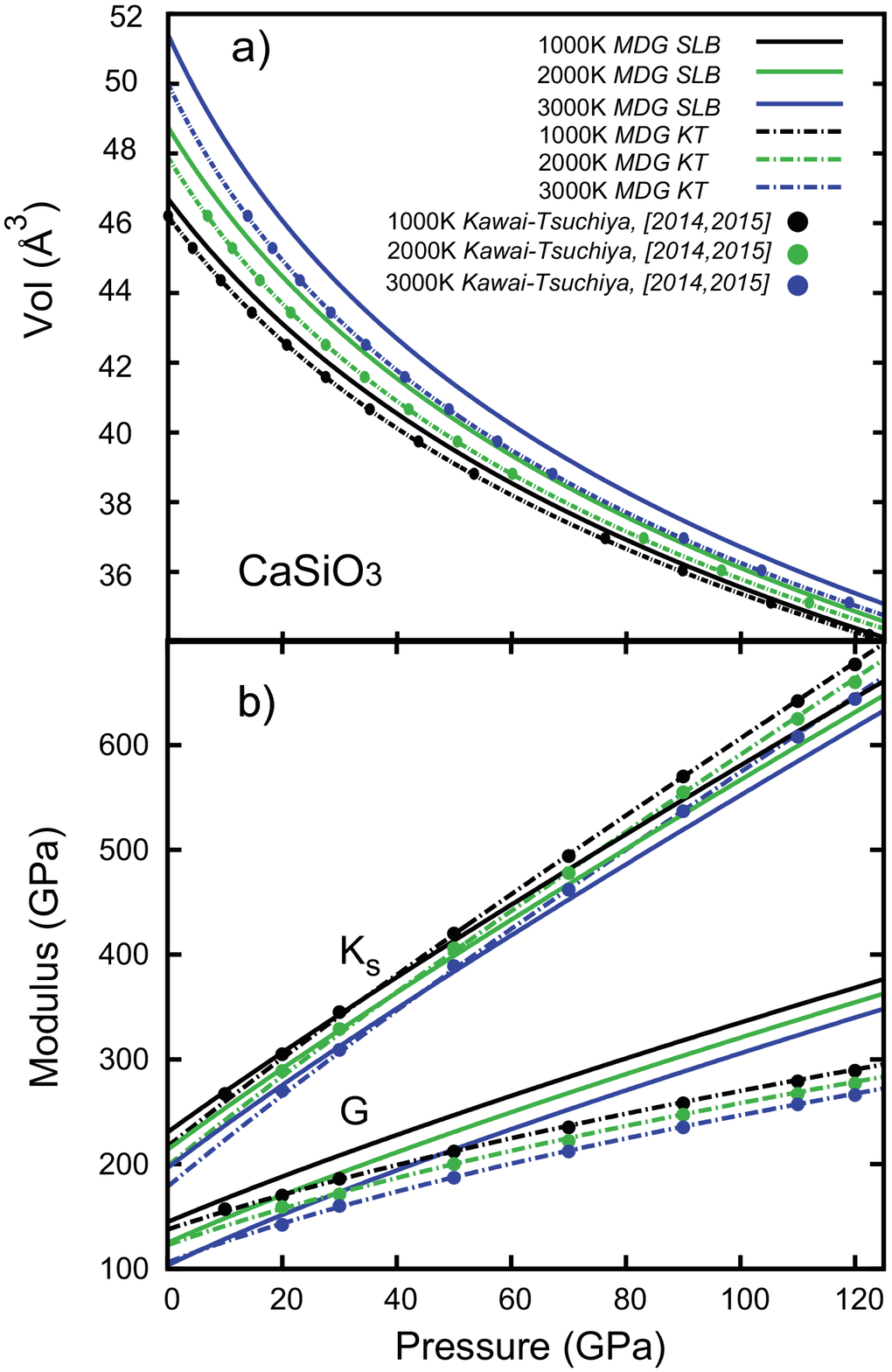}
\caption{CaSiO$_3$ compression curves (a) and moduli (b) for different temperatures. Solid lines represent data obtained using the Mie-Debye-Gr\"uneisen (MDG) model with EoS parameters from \cite{Stixrude2} (SLB). Dashed lines are results calculated using parameters from \cite{Kawai,Tsuchiya} (KW). Dots correspond to direct results reported by \cite{Kawai,Tsuchiya}. }
\label{figure7}
\end{figure}

\end{article}

\begin{landscape}
\begin{table}
\centering
\caption{ Fundamental oxides for lower mantle compositions. Bold data in brackets indicates original values before adjusting Al$_2$O$_3$ moles.\\ 
a \cite{Williams} \\ 
b \cite{Hart} \\ 
c \cite{McDonough} \\ 
d \cite{hirose}  \\ 
e \cite{baker}\\
 ** MgO moles were further modified to avoid excess of this oxide 
}
\label{tbl:oxides}
\resizebox{22cm}{!} {
\begin{tabular}{c c c c c c c c c c c}\hline \\
  	&   Perovskite $^a$ (Mg/Si: 0.82)    &		& Chondrite $^{a,b}$ (Mg/Si: 1.07)   &            &  Pyrolite $^{a,c}$ (Mg/Si: 1.24)  &           &  Peridotite $^d$(Mg/Si: 1.30) &		& Harzburgite $^e$ (Mg/Si: 1.56)\\ 
				& Only  & & & & & & & & & \\ \hline \\
				&  Mol \%     &      Wt \%           &   Mol \%       &      Wt \%  &   Mol \%       &   Wt \%  &   Mol \%     &   Wt \%   &     Mol\% &    Wt \%           \\ \hline \\				  
CaO				&  2.91 \textbf{(2.89)}     &  3.11 \textbf{(3)}  &    2.61 \textbf{(2.65)}    & 2.86 \textbf{(2.82)}           &  3.26 \textbf{(3.33)}      & 3.61 \textbf{(3.55)}  &   3.13 \textbf{(3.19)}   & 3.48 \textbf{(3.44)}             &   0.81 \textbf{(0.81)}  &  0.91 \textbf{(0.91)}        \\
MgO				&  41.34* \textbf{(41.79)}   &  31.76 \textbf{(31.18)} &    47.70 \textbf{(46.74)}   & 37.59 \textbf{(35.68)}   &  50.35 \textbf{(49.3)}     & 40.05 \textbf{(37.8)} &   51.5 \textbf{(50.61)}  & 41.14 \textbf{(39.22)}           &   57.04 \textbf{(56.51)} &  46.17 \textbf{(45.73)}       \\
SiO$_2$				&  50 \textbf{(47.50)}     &  57.26 \textbf{(52.83)} &    44.54 \textbf{(43.52)}  & 52.32 \textbf{(49.52)}    &  40.62 \textbf{(39.37)}    & 48.17 \textbf{(45)}   &   39.61 \textbf{(38.50)} & 47.17 \textbf{(44.48)}           &   36.6 \textbf{(36.07)}  &  44.16 \textbf{(43.51)}       \\
FeO				&  5.74  \textbf{(5.70)}    &  7.86 \textbf{(7.58)}    &    5.15  \textbf{(5.25)}  & 7.24  \textbf{(7.14)}      &  5.77 \textbf{(5.89)}      & 8.18 \textbf{(8.05)}  &   5.76 \textbf{(5.86)}  & 8.2  \textbf{(8.1)}            &   6.04 \textbf{(6.07)}  &  8.76  \textbf{(8.76)}         \\  \hline \\	
Al$_2$O$_3$			&  \textbf{(2.12)}      &   \textbf{(4)}       &    \textbf{(1.84)}    & \textbf{(3.56)}            &  \textbf{(2.11)}     & \textbf{(4.09)}      &   \textbf{(1.83)}   & \textbf{(3.59)}   &   \textbf{(0.53)}   &  \textbf{(1.09)}           \\  \hline\hline	
\end{tabular}
}
\end{table}
\end{landscape}

\newpage

\begin{landscape}
\begin{table}
\caption{ Lower mantle minerals in different aggregates. The Vol \% values are given at \textit{Brown and Shankland}[1981] anchoring conditions.}
\label{tbl:compositions}
\resizebox{22cm}{!} {
\begin{tabular}{lccccccccccccccc}\hline \\
				&    &   Perovskite Only (Mg/Si: 0.82)    &	 &	& Chondritic (Mg/Si: 1.07)   &           &    &  Pyrolite (Mg/Si: 1.24)  &  &   &        Peridotite (Mg/Si: 1.30)  &&&      Harzburgite (Mg/Si: 1.56)    \\ 
				&     &	   $x = 0.1220$ &    &	  & $ x= 0.0829 , y= 0.1532 $& &	& $ x= 0.0725 , y= 0.1353 $&  &  &    $ x= 0.0767 , y= 0.1425 $          &&&     $ x= 0.0698 , y= 0.1306 $&        \\ \hline \\
				&  Mol \% &  Vol \%  &   Wt \% $\qquad$  &  Mol \%   &  Vol \%      &   Wt \% $\qquad$  &   Mol \%    &  Vol \%       &    Wt \% $\qquad$ &   Mol \%   &  Vol \% &   Wt \%  $\qquad$  &  Mol \%  & Vol\%    &      Wt \% 	  \\ \hline \\				   
(Mg$_{1-x}$ Fe$_x$)SiO$_3$	& 94.18   &  93.54   &  93.55  $\qquad$     & 75.61      &  84.16   &	84.44 $\qquad$   &  65.51      &  77.84       & 78.37  $\qquad$   &   60.4    & 73.8     & 74.33 $\qquad$     &  55.99     & 72.87   & 73.74      \\
(Mg$_{1-y}$Fe$_y$)O	        &  0      &  0       &   0     $\qquad$     & 19.69      &  9.97    &	9.64  $\qquad$  &  30.03      &  16.23        & 15.59  $\qquad$   &   34.41   & 19.11    & 18.45 $\qquad$     &  42.74     & 25.28   & 24.37      \\
CaSiO$_3$ 		        & 5.82    &  6.46    &   6.45  $\qquad$     & 4.7        &  5.85    &	5.92  $\qquad$  &  4.46       &   5.93        &  6.04  $\qquad$   &   5.19    & 7.09     & 7.22  $\qquad$     &  1.27      &  1.85   & 1.89       \\ \hline\hline		
\end{tabular}                                                                                                                                                                                            
}
\end{table}
\end{landscape}

\newpage

\begin{landscape}
\begin{table}
\caption{Adiabats for lower mantle minerals. $x = 0.125$ , $y = 1875$.}
\label{tbl:compositions}
\resizebox{22cm}{!} {
\begin{tabular}{lccccccccccccccc}\hline \\
				&    &   Mg$_{1-y}$Fe$_{y}$O HS    &	 &	& Mg$_{1-y}$Fe$_{y}$O MS    &           &    &  Mg$_{1-x}$Fe$_{x}$SiO$_3$  &  &   &        CaSiO$_3$   \\ 
$\qquad$   Pressure  (GPa)    	&    &	 Temperature (K) &    &	  &  Temperature (K) & &	&  Temperature (K) &  &  &    Temperature  (K)                \\ \hline \\
$\qquad$ $\qquad$ 23		&    &  1873.00      &    $\qquad$     &       &  1873.00    &	 $\qquad$  &     &  1873.00        &   $\qquad$   &     & 1873.00    &       \\
$\qquad$ $\qquad$ 30	        &    &  1945.85      &    $\qquad$     &       &  1945.85    &	 $\qquad$  &     &  1927.92        &   $\qquad$   &     & 1929.92    &       \\
$\qquad$ $\qquad$ 40	        &    &  2050.38      &    $\qquad$     &       &  2050.38    &	 $\qquad$  &     &  2006.76        &   $\qquad$   &     & 2011.78    &       \\
$\qquad$ $\qquad$ 50	        &    &  2144.59      &    $\qquad$     &       &  2144.59    &	 $\qquad$  &     &  2078.87        &   $\qquad$   &     & 2086.67    &       \\
$\qquad$ $\qquad$ 60	        &    &  2230.87      &    $\qquad$     &       &  2238.41    &	 $\qquad$  &     &  2145.39        &   $\qquad$   &     & 2155.80    &       \\
$\qquad$ $\qquad$ 70	        &    &  2310.84      &    $\qquad$     &       &  2328.82    &	 $\qquad$  &     &  2207.19        &   $\qquad$   &     & 2220.08    &       \\
$\qquad$ $\qquad$ 80	        &    &  2385.71      &    $\qquad$     &       &  2438.50    &	 $\qquad$  &     &  2264.98        &   $\qquad$   &     & 2280.21    &       \\
$\qquad$ $\qquad$ 90	        &    &  2456.34      &    $\qquad$     &       &  2556.66    &	 $\qquad$  &     &  2319.33        &   $\qquad$   &     & 2336.73    &       \\
$\qquad$ $\qquad$ 100	        &    &  2523.40      &    $\qquad$     &       &  2672.30    &	 $\qquad$  &     &  2370.71        &   $\qquad$   &     & 2390.09    &       \\
$\qquad$ $\qquad$ 110	        &    &  2587.40      &    $\qquad$     &       &  2782.20    &	 $\qquad$  &     &  2419.51        &   $\qquad$   &     & 2440.64    &       \\
$\qquad$ $\qquad$ 120	        &    &  2648.73      &    $\qquad$     &       &  2884.95    &	 $\qquad$  &     &  2466.06        &   $\qquad$   &     & 2488.70    &       \\
$\qquad$ $\qquad$ 125		&    &  2678.49      &    $\qquad$     &       &  2932.98    &	 $\qquad$  &     &  2488.58        &   $\qquad$   &     & 2511.87    &       \\ \hline\hline		
\end{tabular}                                                                                                                                                                            
}
\end{table}

\begin{table}
\caption{Adiabats for lower mantle aggregates with fp in MS state.}
\label{tbl:compositions}
\resizebox{22cm}{!} {
\begin{tabular}{lccccccccccccccc}\hline \\
				&    &   Perovskite Only    &	 &	& Chondritic   &           &    &  Pyrolite  &  &   &        Peridotite   &&&      Harzburgite    \\ 
$\qquad$   Pressure  (GPa)    	&     &	 Temperature (K) &    &	  &  Temperature (K) & &	&  Temperature (K) &  &  &    Temperature  (K)         &&&     Temperature (K) &        \\ \hline \\
$\qquad$ $\qquad$ 23		&    &  1873.00      &    $\qquad$     &       &  1873.00    &	 $\qquad$  &     &  1873.00        &   $\qquad$   &     & 1873.00    &  $\qquad$     &       & 1873.00   &      \\
$\qquad$ $\qquad$ 30	        &    &  1928.04      &    $\qquad$     &       &  1929.67    &	 $\qquad$  &     &  1930.64        &   $\qquad$   &     & 1931.13    &  $\qquad$     &       & 1931.97   &      \\
$\qquad$ $\qquad$ 40	        &    &  2007.05      &    $\qquad$     &       &  2010.99    &	 $\qquad$  &     &  2013.33        &   $\qquad$   &     & 2014.52    &  $\qquad$     &       & 2016.54   &      \\
$\qquad$ $\qquad$ 50	        &    &  2079.32      &    $\qquad$     &       &  2085.24    &	 $\qquad$  &     &  2088.76        &   $\qquad$   &     & 2090.56    &  $\qquad$     &       & 2093.59   &      \\
$\qquad$ $\qquad$ 60	        &    &  2149.18      &    $\qquad$     &       &  2153.92    &	 $\qquad$  &     &  2158.59        &   $\qquad$   &     & 2161.02    &  $\qquad$     &       & 2165.02   &      \\
$\qquad$ $\qquad$ 70	        &    &  2207.92      &    $\qquad$     &       &  2218.68    &	 $\qquad$  &     &  2224.83        &   $\qquad$   &     & 2228.17    &  $\qquad$     &       & 2233.42   &      \\
$\qquad$ $\qquad$ 80	        &    &  2265.85      &    $\qquad$     &       &  2281.19    &	 $\qquad$  &     &  2289.55        &   $\qquad$   &     & 2294.41    &  $\qquad$     &       & 2301.50   &      \\
$\qquad$ $\qquad$ 90	        &    &  2320.32      &    $\qquad$     &       &  2341.75    &	 $\qquad$  &     &  2352.92        &   $\qquad$   &     & 2359.78    &  $\qquad$     &       & 2369.19   &      \\
$\qquad$ $\qquad$ 100	        &    &  2371.81      &    $\qquad$     &       &  2399.31    &	 $\qquad$  &     &  2413.30        &   $\qquad$   &     & 2422.18    &  $\qquad$     &       & 2433.94   &      \\
$\qquad$ $\qquad$ 110	        &    &  2420.71      &    $\qquad$     &       &  2453.19    &	 $\qquad$  &     &  2469.59        &   $\qquad$   &     & 2480.21    &  $\qquad$     &       & 2494.04   &      \\
$\qquad$ $\qquad$ 120	        &    &  2467.35      &    $\qquad$     &       &  2503.53    &	 $\qquad$  &     &  2521.81        &   $\qquad$   &     & 2533.77    &  $\qquad$     &       & 2549.28   &      \\
$\qquad$ $\qquad$ 125		&    &  2489.90      &    $\qquad$     &       &  2527.55    &	 $\qquad$  &     &  2546.59        &   $\qquad$   &     & 2559.08    &  $\qquad$     &       & 2575.28   &      \\ \hline\hline		
\end{tabular}                                                                                                                                                                            
}
\end{table}

\end{landscape}

\end{document}